\documentclass[10pt, fleqn]{article}
\usepackage[utf8]{inputenc}
\DeclareUnicodeCharacter{2212}{$-$}
\usepackage{graphicx}
\usepackage[fleqn]{amsmath}
\usepackage[version=4]{mhchem}
\usepackage{siunitx}
\usepackage{longtable,tabularx}
\usepackage{nccmath}
\usepackage{amsmath}
\usepackage[superscript]{cite}
\usepackage[margin=1in]{geometry}
\usepackage[toc,page]{appendix}
\usepackage{setspace}
\usepackage{amssymb}
\usepackage{tabu}
\usepackage[colorlinks,urlcolor=blue,linkcolor=blue,citecolor=blue]{hyperref}
\usepackage{afterpage}
\usepackage{titlesec}
\usepackage[labelfont=bf]{caption}
\usepackage{fancyhdr}
\usepackage{pgf}
\usepackage{float}
\usepackage{subcaption}
\usepackage{cleveref}
\usepackage{textcomp}
\usepackage{pict2e}
\usepackage{wasysym}
\usepackage{multicol}
\usepackage{lipsum}
\usepackage{indentfirst}
\newenvironment{Figure}
  {\par\medskip\noindent\minipage{\linewidth}}
  {\endminipage\par\medskip}

\definecolor{mygreen}{rgb}{0,0.6,0}
\definecolor{mygray}{rgb}{0.5,0.5,0.5}
\definecolor{mymauve}{rgb}{0.58,0,0.82}

\usepackage{listings}
\titleformat{\section}
{\normalfont\normalsize\bfseries}{\thesection}{0.5em}{}
\titleformat{\subsection}
{\normalfont\normalsize\bfseries\em}{\thesubsection}{0.5em}{}
\titleformat{\subsubsection}
{\normalfont\normalsize\bfseries\em}{\thesubsubsection}{0.5em}{}
\begin{document}

\begin{flushright}
\Large % (14 point font)

% Input the paper number: 
\textbf{[SSC22-XII-04]}
\end{flushright}
\begin{centering}      
\large % (12 point font)

% Input the title: 
\textbf{First demonstration of a post-quantum key-exchange with a nanosatellite}\\
\vspace{0.5cm}
\normalsize % set to 10 point font

% Input the author information:
{Simon M. Burkhardt, Willi Meier, Christoph F. Wildfeuer}\\
{Institute for Sensors and Electronics, University of Applied Sciences and Arts Northwestern Switzerland}\\
{Klosterzelgstr. 2, 5210 Windisch}; +41-562028792\\
{simon.burkhardt@fhnw.ch}\\ 
\vspace{0.5cm}
{Ayesha Reezwana, Tanvirul Islam, Alexander Ling}\\
{Centre for Quantum Technologies, National University of Singapore}\\
{3 Science Drive 2, 120435}; +65-86545795\\
{cqtayes@nus.edu.sg}\\

% Input the abstract: 
\vspace{0.5cm}
\centerline{\textbf{ABSTRACT}}
\vspace{0.3cm}
\end{centering}

We demonstrate a post-quantum key-exchange with the nanosatellite SpooQy-1 in low Earth orbit using Kyber-512, a lattice-based key-encapsulation mechanism and a round three finalist in the NIST PQC standardization process. Our firmware solution runs on an on-board computer that is based on the Atmel AVR32 RISC microcontroller, a widely used platform for nanosatellites. We uploaded the new firmware with a 436.2 MHz UHF link using the CubeSat Space Protocol (CSP) and performed the steps of the key exchange in several passes over Switzerland. The shared secret key generated in this experiment could potentially be used to encrypt RF links with AES-256. This implementation demonstrates the feasibility of a quantum-safe authenticated key-exchange and encryption system on SWaP constrained nanosatellites.
   
\begin{multicols*}{2}

\section*{Introduction} 
Security techniques for commercial satellites are poorly developed despite the rapid increase in the number of satellite missions. New constellations will increase the number of satellites dramatically over the next decade. It can be expected that with an increase in the number of operational satellites, the number of cyber-attacks on spacecraft communication will also increase. \cite{Manulis}

\begin{Figure}
    \centering
    \includegraphics[width=17.5pc]{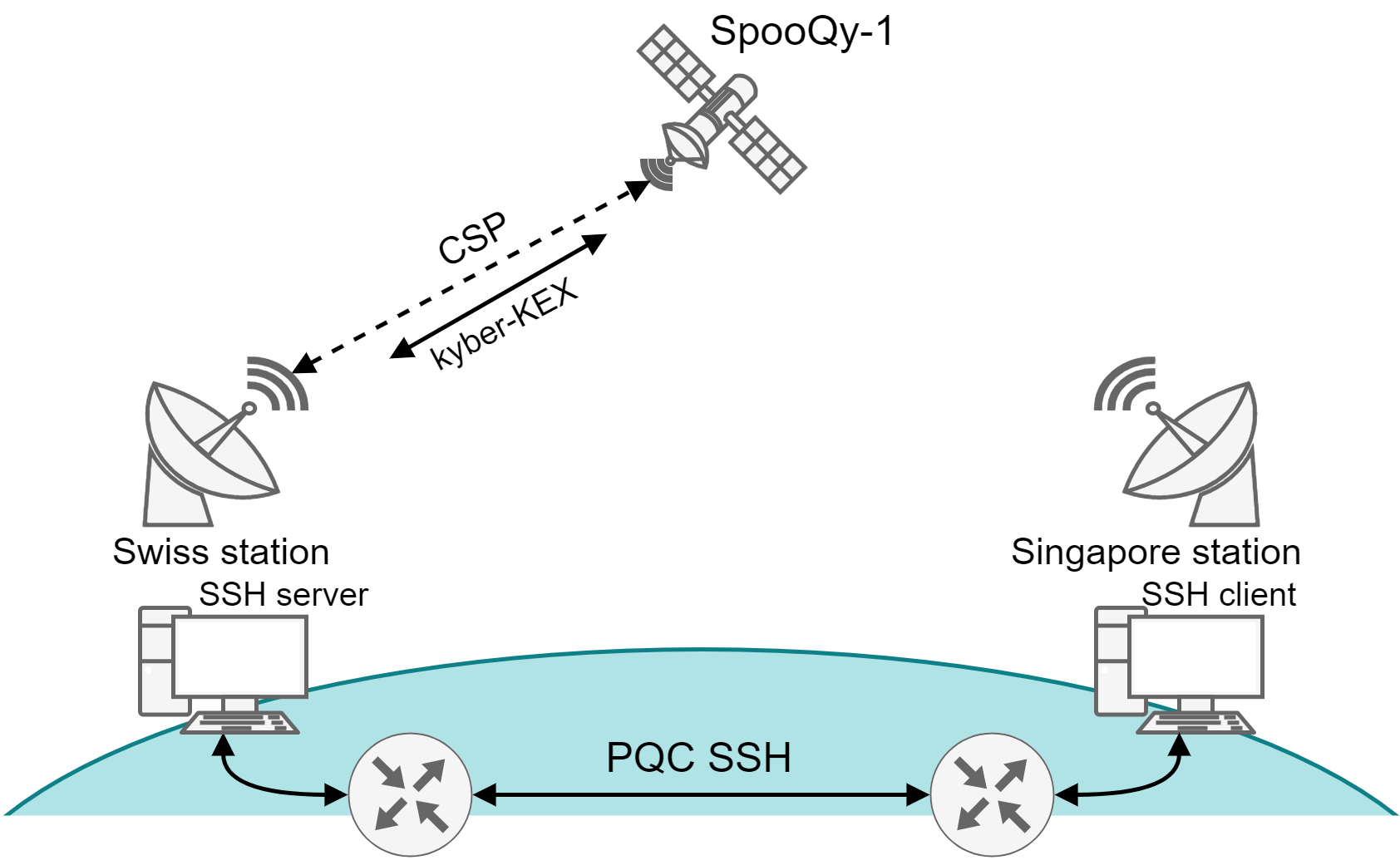}
	\captionof{figure}{Satellite infrastructure used in this experiment}
	\label{fig:titlefig}
\end{Figure}

\par In SatCom systems, the symmetric key encryption algorithms - where the same key is used for both encoding and decoding messages - are frequently used. This approach to encryption in satellite communication requires that every party establishes a unique secret key with every other party with whom they would like to communicate. Furthermore, adding to this problem, each party must obtain all its secret keys in advance because possession of an appropriate key is a necessary prerequisite to establish a secure communication channel with another party. The number of shared key pairs that every party needs to store increases according to $n(n-1)/2$, where $n$ may be the number of satellites in a constellation. Considering 100 satellites, this totals 4950 key pairs. Key-management can therefore become an important challenge for larger satellite fleets.
To solve these issues, an asymmetric key encryption algorithm could be adopted where the key used to encrypt the message is different from the key used to decrypt the message. Such an approach requires each of the communication parties to maintain two keys only - one that is kept private and a second one that is made publicly available. 

\par Using public-key exchange protocols for space applications has recently been proposed \cite{PKI_for_Space}. In most public-key systems the public keys are generated with RSA and Elliptic Curve Cryptography (ECC). However, it is now anticipated that Quantum Computers (QC) will be able to break both RSA and ECC when the technology to manufacture enough quantum nodes becomes available.

\par In order to solve this problem, the National Institute of Standards and Technology (NIST) has initiated a process to solicit, evaluate, and standardize one or more quantum-resistant public-key cryptographic algorithms. The goal of Post-Quantum Cryptography (PQC), (also called quantum-resistant cryptography) is to develop cryptographic systems that are secure against both quantum and classical computers and can interoperate with existing communication protocols and networks. Kyber is one of the finalists in the NIST PQC project and it was chosen because it is a secure and efficient Key Encapsulation Mechanism (KEM), whose security is based on the hardness to solve the Learning-with-Errors (LWE) problem over module lattices. \cite{KYBER, NIST}.

\par Our SpooQy-1 satellite consisted of two GomSpace components which handle communication, commands and data. The AX100 COM module had a SHA-1 Hash-based Message Authentication scheme (HMAC) but no encryption for the data. SHA-1 is well known for being insecure \cite{SHA1Collision}.

\par The second module, the A3200 On-Board Computer (OBC) was used for flight controls and mission software. The OBC ran the open source libcsp \cite{LIBCSP_GITHUB} implementation of the CSP which offers optional encryption using a 128-bit symmetric XTEA algorithm.
% which offered none or optionally, very weak encryption methods 
%\textcolor{red}{Can we write this? 
%The OBC ran the open source libcsp \cite{LIBCSP_GITHUB} implementation of the CubeSat Space Protocol (CSP) which offers no security and additionally possesses very weak encryption methods}
%\textcolor{green}{(simon) yes: it is open source on github. about the encryption: we could clarify that it is XTEA and leave away the comment about "very weak" (see above)} 
With the goal of moving towards a quantum secure satellite infrastructure (including CubeSats), it is crucial to embed PQC into current hardware and software projects. Many of the CubeSats run on Size, Weight and Power (SWaP) constrained on-board computers. We present a demonstration of a successful key exchange with this experiment using the older AVR32 microcontroller architecture as well as some performance measurements on the more recent ARM Cortex-M4 architecture. Figure \ref{fig:titlefig} shows our setup of the ground stations and the SpooQy-1 satellite.

\section*{SpooQy-1 CubeSat}
\subsection*{Development and objectives}
\par The main objective of SpooQySat, the SpooQy-1 CubeSat, was to demonstrate an in-orbit space-compatible quantum light source SPEQS (the Small Photon Entangling Quantum System)  to increase the Technology Readiness Level (TRL) of future global Quantum Key Distribution (QKD) networks. QKD is a family of secure communication techniques used to generate private and shareable random secret keys that can be exchanged between two parties only. Essentially, QKD requires the exchange of individual photons and therefore very low-loss optical links need to be established. Optical fibers are limited to about 100 km before losses become overwhelming. Free-space optical losses are much lower. However, the main drawback for free-space QKD is key exhaustion due to failed key generation because of bad weather. Here PQC can provide a fallback solution if keys cannot be exchanged by QKD. Also, PQC may become the standard for encrypting the RF wireless satellite data links since it does not require special hardware and optical communication. That was the motivation to implement a PQC algorithm on SpooQy-1 after the main objective for the mission had been accomplished.
\begin{Figure}
    \centering
    \includegraphics[width=17.5pc]{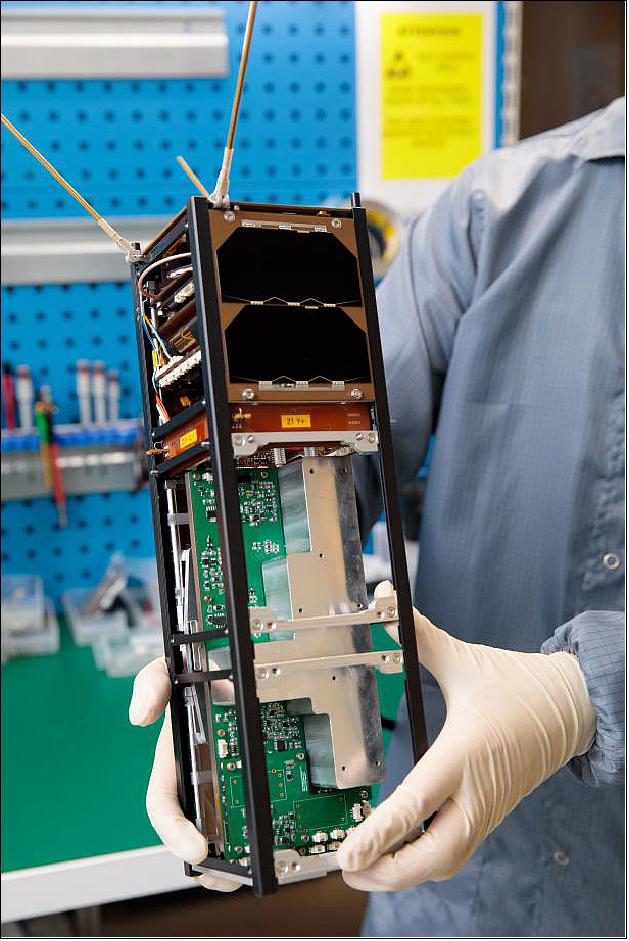}
	\captionof{figure}{Partially integrated engineering model of SpooQySat, a 3U CubeSat. Removed solar panels reveal structural model of the SPEQS payload.}
	\label{fig:spooqy}
\end{Figure}

%\par Experiments started already in 2012 with high-altitude balloon tests of a basic SPEQS source followed by a correlated SPEQS source in 2013  \cite{HAP_1,HAP_2}
\par Experiments of a basic SPEQS source started in 2012 with high-altitude balloon tests followed by a correlated SPEQS source in 2013 \cite{HAP_1,HAP_2}. In 2016, a Space-qualified, correlated SPEQS source was tested in low Earth orbit on the NUS Galassia CubeSat \cite{Galassia}.

\iffalse
\begin{figure}
    \centering
    \subfloat[label 1]{{\includegraphics[width=17.5pc]{sg_ground_station.jpg} }}
    \caption{Singapore UHF ground station on the roof top at NUS campus.}
    \qquad
    \subfloat[label 2]{{\includegraphics[width=17.5pc]{ch_groundstation.jpg} }}
    \caption{2 Figures side by side}
    \label{fig:example}
\end{figure}
\fi

\iffalse
\documentclass{article}
\usepackage[demo]{graphicx} % remove 'demo' option for production version of file
\begin{document}

\begin{figure}[h!]
\begin{minipage}[t]{0.48\textwidth}
\includegraphics[width=\linewidth,keepaspectratio=true]{sg_ground_station.jpg}
\caption{Figura experimental}
\label{fase1}
\end{minipage}
\hspace*{\fill} % it's important not to leave blank lines before and after this command
\begin{minipage}[t]{0.48\textwidth}
\includegraphics[width=\linewidth,keepaspectratio=true]{f}
\caption{Altra figura experimental}
\label{fase2}
\end{minipage}
\end{figure}
\end{document}
\fi

\begin{Figure}
    \centering
    \includegraphics[width=17.5pc]{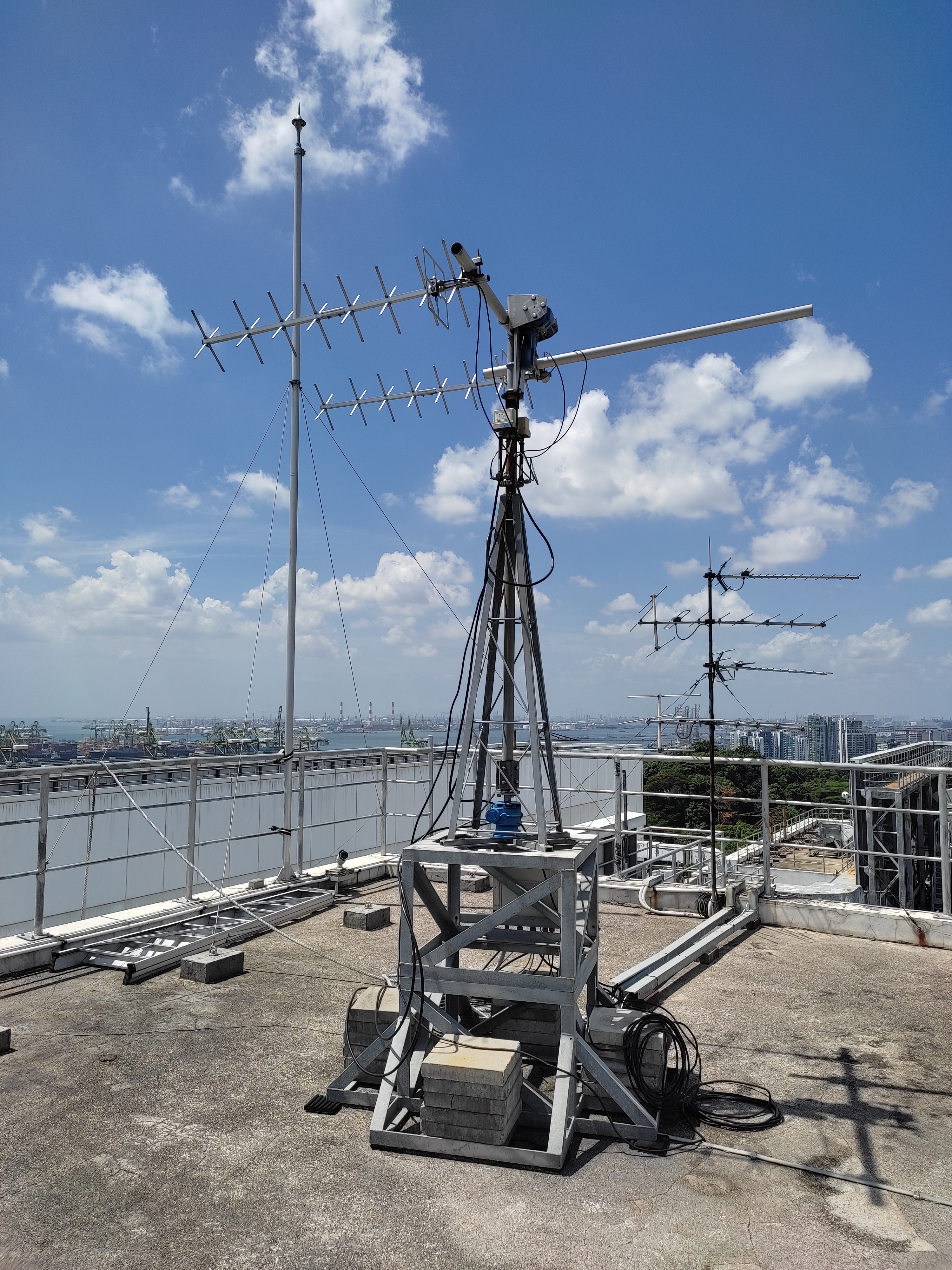}
	\captionof{figure}{Singapore UHF ground station on the roof top at NUS campus.}
	\label{fig:sg_gs}
\end{Figure}

\begin{Figure}
    \centering
    \includegraphics[width=17.5pc]{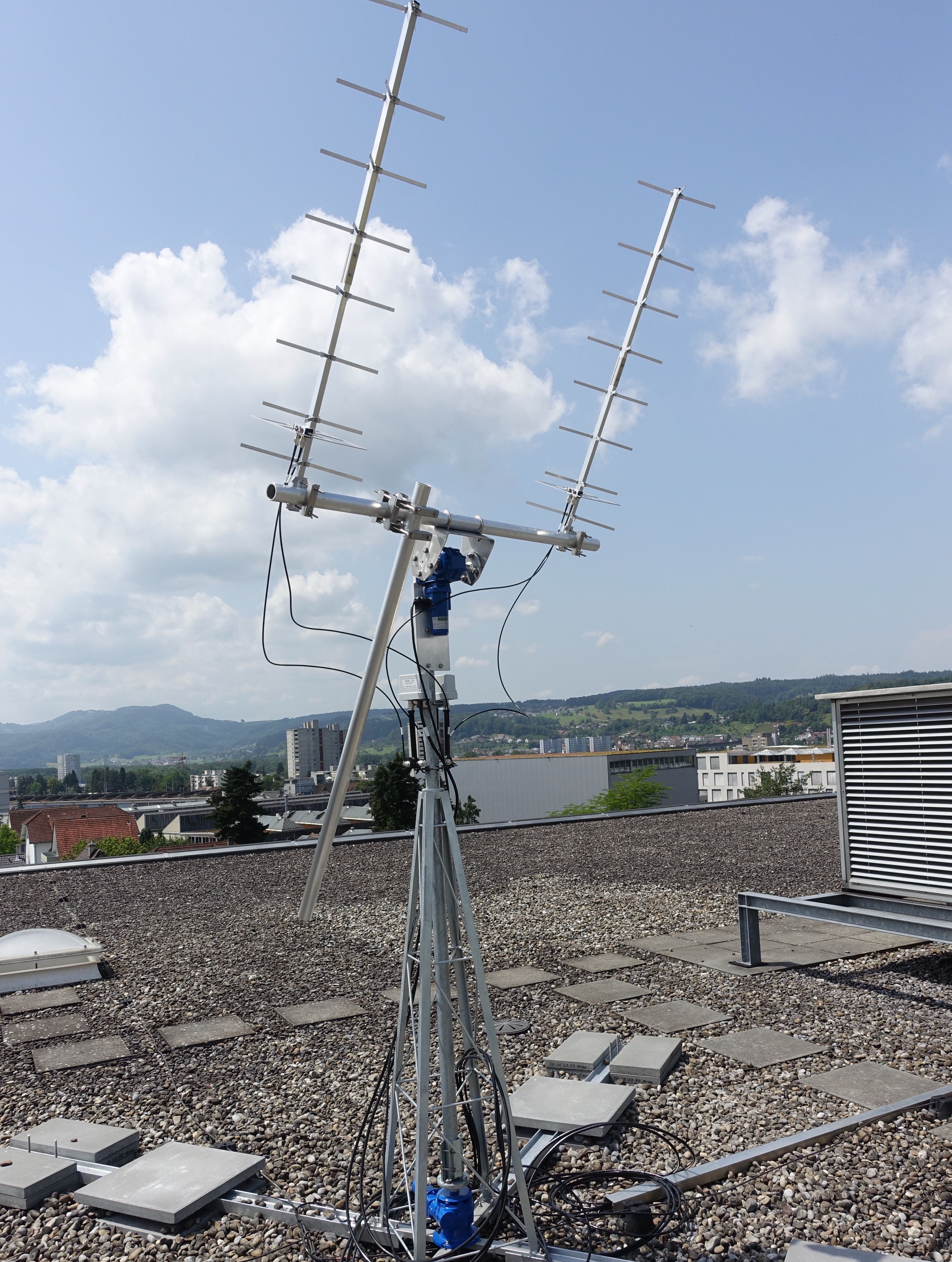}
	\captionof{figure}{Switzerland UHF ground station at the campus of FHNW in Windisch.}
	\label{fig:ch_gs}
\end{Figure}

\par SpooQy-1 was then designed and built at the Centre for Quantum Technologies, National University of Singapore to demonstrate an entangled photon pair-source in space. SpooQy-1 was deployed to Low Earth Orbit (LEO) from the international space station on 17th June 2019 and provided the first demonstration of entanglement in space on a nanosatellite \cite{SPOOQY1}. In Figure \ref{fig:spooqy} the partially integrated engineering model is shown. Fully assembled, the CubeSat mass is 2.6 kg, and its peak system power consumption is 3.9W.

\par The Singapore ground station is located on top of an eighteen storied building at the NUS campus shown in Figure \ref{fig:sg_gs}. A secondary UHF ground station, shown in Figure \ref{fig:ch_gs}, is established in Switzerland to provide additional data download opportunities. The ground stations are built using the GomSpace UHF hardware and have identical setups. Both ground stations are equipped with a twinned Yagi antenna with a tracking mount. The rotor is controlled by a Linux based server computer (NanoCom MS100). The ground station radio (NanoCom GS100) is the ground counterpart (with a 25 W power amplifier) for the NanoCom AX100 radio on-board SpooQy-1, designed specifically as an integrated component to request/respond via the CSP protocol during operation.

\subsection*{AVR32 platform}
The SpooQy-1 nanosatellite uses the NanoMind A3200 on-board computer from GomSpace which utilizes a Microchip AT32UC3C0512C micro controller running a real time operating system (FreeRTOS) along with proprietary mission specific software. On-board are 128MB of external flash storage which can be accessed through the C stdlib file IO functions. On the flash memory there is a FAT file system present which can be accessed through an FTP implementation for CSP. An additional 32MB of SDRAM can be used to load a binary RAM image file \texttt{nanomind.elf} from the file system and boot from there. This enables the satellite with the capability to run new code once it is in orbit.

\subsection*{Firmware framework}

GomSpace delivers the NanoMind with a Software Development Kit (SDK) and documentation to build and expand mission firmware for their AVR platform. The SDK consists of a fully featured mission control software with the software parts.
%as listed in Table \ref{tab:sdk}. 
The Figure \ref{fig:hwsw_diagram} shows a visual representation of the software components that are involved both in the ground station and the satellite. 

%\begin{center}
%\fontsize{4}{6}\selectfont
%\captionof{table}{Software parts of the SDK}
%\begin{table}[]
%\resizebox{200}{!}{
%    \begin{tabular}{|l|}
%    \hline
%    Utilities and drivers         \T\B \\ \hline
%    FreeRTOS 8.0.1                \T\B \\ \hline
%    GomSpace Shell (GOSH)         \T\B \\ \hline
%    CSP                           \T\B \\ \hline
%    Parameter System              \T\B \\ \hline
%    File Transfer Protocol        \T\B \\ \hline
%    Housekeeping (Telemetry)      \T\B \\ \hline
%    Flight Planner                \T\B \\ \hline
%    \end{tabular}
%}
%\end{table}
%\label{tab:sdk}
%\end{center}

%\begin{itemize}
%    \item Utilities (GOSH) and drivers
%   \item FreeRTOS 8.0.1
%    \item CSP
%    \item Parameter System
%    \item File Transfer Protocol
%    \item Housekeeping (Telemetry)
%    \item Flight Planner
%\end{itemize}

Our goal was to implement the Kyber algorithm alongside this mission control software to demonstrate that the Kyber source code can run on the AVR32 platform. The following chapter describes the challenges, solutions and recommendations when using PQC algorithms on satellite hard- and firmware.

\section*{Implementing the key exchange}
\subsection*{Kyber key encapsulation mechanism}
The PQ CRYSTALS Kyber algorithm is a quantum secure Key Encapsulation Mechanism (KEM). A KEM can be used in combination with a Key Derivation Function (KDF) to generate a common symmetric key. In the case of Kyber, SHA-256 is used as the KDF. The reference implementation from PQ CRYSTALS contains the API source code for such a KEM inside \texttt{kex.c} as well as a principal protocol definition in section 5 of \cite{KYBER}. The API offers two types of key exchange: the “Unilaterally Authenticated Key Exchange” (UAKE) and the “Mutually Authenticated Key Exchange” (AKE), which is the preferred and most secure method. Whereby the authentication does not authenticate the participants but guarantees that each party has derived the same symmetric key. For user authentication a separate algorithm like Dilithium \cite{DILITHIUM} would be required. In our experiment we had the advantage of using pre-shared secrets to authenticate both parties using HMAC that had already been implemented in the AX100 COM module. However, since the implemented HMAC is based on SHA-1 it is not quantum-safe.

\subsection*{AVR32 toolchain}
Compilation of the GomSpace firmware is done using the AVR32 tool chain (version 3.4.2). It contains \texttt{avr32-gcc} (gcc version 4.4.7) for Debian-based Linux systems. A drawback of this outdated C compiler is that it only supports C language up to the C99 standard. The Kyber implementation is included in the liboqs project from the open quantum safe organization \cite{LIBOQS_GITHUB}. Liboqs is a sandbox to experiment with many different PQC algorithms which are participating in the NIST standardization process. However, this project requires C11 standard and utilizes functions like \texttt{aligned\_alloc} which are unavailable from the \texttt{avr32-gcc} compiler. We therefore focused on the standalone Kyber algorithm rather than including multiple NIST candidate algorithms. In a first step, the original Kyber source code from the PQ CRYSTALS organization \cite{KYBER_GITHUB} was integrated into the AVR32 auto build system for the NanoMind. This step includes platform specific adjustments to the Kyber source code like the random number source.

\subsection*{Random number generators}
To guarantee quantum safety, the Kyber algorithm requires a Random Number Generator (RNG) that can produce 256 bits of entropy. The original Kyber implementation uses the \texttt{/dev/urandom} pseudo-random number generator on Linux based systems. Although being pseudo-random, it is considered to be safe for cryptographic applications. \cite{URAND1, URAND2}
\par SpooQy-1's OBC is only running an RTOS and not a full operating system which would offer such a secure RNG. This is why the Kyber source code was modified to replace reading from \texttt{/dev/urandom} with the pseudo-random function \texttt{rand()} from the \texttt{avr32-gcc} stdlib.  
This affects the \texttt{randombytes()} function used during the asymmetric key pair generation. Another issue is that on AVR32 a True Random Number Generator (TRNG) is missing. For time reasons we decided not to implement SpooQy-1's on-board Quantum Random Number Generator (QRNG) into this experiment. In a practical application the \texttt{rand()}-function is the weakest point of failure because the private keys from pseudo-randomly generated key pairs can sometimes be recovered as demonstrated in \cite{RNG1}. We strongly advise, using a TRNG or even a QRNG to get the required 256 bits of entropy for Kyber. As a less secure alternative one can seed the PRNG using a true random number or a pre-shared secret seed using \texttt{srand(seed)}. In this experiment we used the default seed. It should also be pointed out that the ground station is using the secure RNG, as depicted in Figure \ref{fig:kex_diagram}. If the ground station initiates the key exchange, the random ingredients for the common secret are therefore cryptographically secure. This would not be the case, if the satellite initiates the key exchange.

\subsection*{Practical key exchange application}
The original implementation in \texttt{test\_kex.c} from the Kyber source code served as a reference implementation for our AVR32 application \cite{KYBER_GITHUB}. Both the SDK for the satellite and the ground station allow the developer to implement callback handlers for custom features on both ends (in our source code: \texttt{kex\_pub.c} and \texttt{kex\_kyber.c}, respectively \cite{FHNW_GITHUB}). We implemented the Kyber API into callback functions for the satellite's command parser. Limited by the available time to develop such an implementation, we decided to implement only the satellite's back end and perform the message exchange manually using the File Transfer Protocol (FTP). All keys and temporary arrays are not just stored in RAM but also in hex-format as ASCII characters on the NanoMind’s and ground station's file systems. This has another practical reason: in case of a reboot the keys can be recovered from the flash memory. During the key exchange the encapsulated message files are not exchanged automatically but manually using FTP upload and download commands. For this purpose, we implemented a way to read and write these message files on both stations. The two code snippets in Listing \ref{lst:writeFile} and Listing \ref{lst:sharedB} show the principle behind the code that was executed as part of our key exchange experiment. The full code is available in our Github repository \cite{FHNW_GITHUB}.

\begin{lstlisting}[caption={Source code of how the messages\\ are written to files inside\\ writeHexFile(filename,in,bytes).},label=lst:writeFile]
  FILE *fp;
  fp = fopen(filename, "w+");
  for(int i=0;i<(int)bytes;i++)
    fprintf(fp,"%02x", in[i]);
  fclose(fp);
\end{lstlisting}

\begin{lstlisting}[caption={Source code of how\\ kex\_ake\_sharedB() is performed with\\ ake\_senda.txt as input.},label=lst:sharedB]
  readHexFile("/flash/ake_senda.txt", ake_senda, KEX_AKE_SENDABYTES);
  kex_ake_sharedB(ake_sendb, ka, ake_senda, ska, pkb); // Run by Bob
  writeHexFile("/flash/COMMON.key", ka, KEX_SSBYTES); // final key
  writeHexFile("/flash/ake_sendb.txt", ake_sendb, KEX_AKE_SENDBBYTES);
\end{lstlisting}

The file “COMMON.key” then contains the HEX representation of the exchanged key. 
This key is 32 bytes long, or 64 characters if stored in ASCII HEX format using \texttt{printf(\%02x)}.

\begin{lstlisting}[caption={The contents of COMMON.key after the successful key exchange.}]
  f02473a6ab18617b3e0dbcc565b4b64e23 f12a284a6dbfbf5cd3cde4ac5e2e21
\end{lstlisting}

\subsection*{Communication channel}

\begin{Figure}
    \centering
    \includegraphics[width=12.5pc]{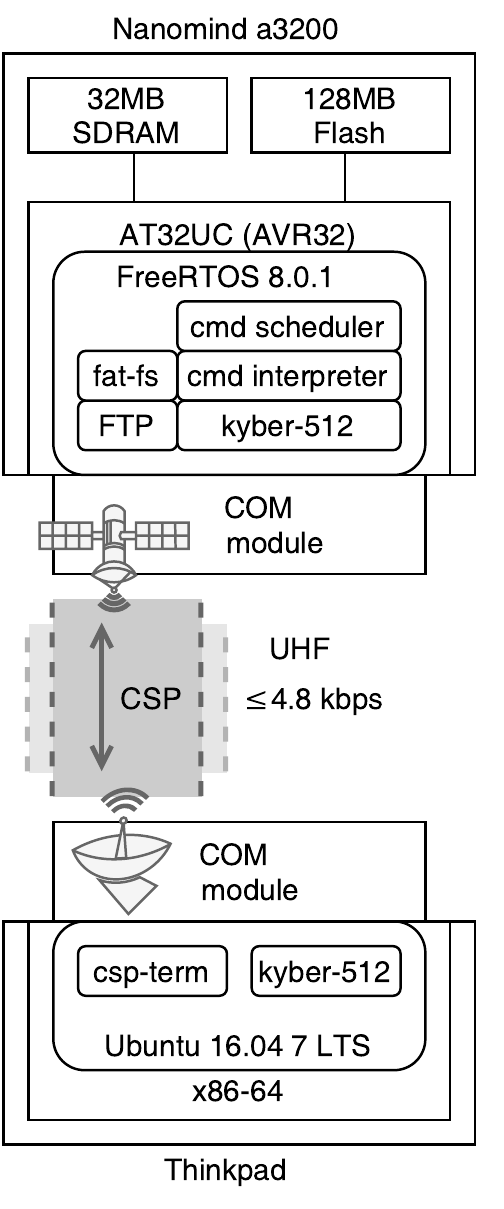}
	\captionof{figure}{Hardware and software components involved in the setup.}
	\label{fig:hwsw_diagram}
\end{Figure}

\subsection*{Resource utilization}
As shown in Table \ref{tab:utilization} our application of Kyber-512 uses approximately 13 kB more flash (\texttt{text}) and 8 kB more RAM (\texttt{bss}) than the default mission firmware without Kyber. The additional 8032 bytes in the RAM are a direct consequence of the several global arrays required to store the keys and messages. 
This is an increase of 40x (flash) respectively 500x (RAM) compared to the insecure XTEA implementation which uses a 128-bit key. If compiled as a RAM image to boot from the SDRAM, less RAM is available for the program which limits the RAM usage of the mission firmware. This is especially relevant for the large arrays used to hold the keys. In our case there is only enough RAM to have temporary arrays for one key exchange. The on-board simulation of a key exchange using two parties would not fit into the memory.

\begin{center}
\fontsize{8}{7}\selectfont
\captionof{table}{Memory usage (in bytes) on AVR32 compared to the default project without any cryptographic function, when compiling for a RAM image.}
%\begin{table}[]
	\small
    \begin{tabular}{|l|r|r|}
    \hline \textbf{Algorithm} & \textbf{flash} & \textbf{RAM} \\ \hline
    default (none) & 467'608   & 31'024     \\ \hline
    XTEA       &   +344        &        +16 \\ \hline
    SHA1-HMAC  & +1'856        &        +16 \\ \hline
    XTEA \& SHA1-HMAC & +2'192      &        +32 \\ \hline
    Kyber-512  & +12'976       & +8'032     \\ \hline
    Kyber-718  & +13'016       & +11'680    \\ \hline
    Kyber-1024 & +13'248       & +15'808    \\ \hline
    \end{tabular}
%\end{table}
\label{tab:utilization}
\end{center}

\section*{Exchanging keys}
\subsection*{Setup}
For the demonstration we assume that the ground station is “Alice” and SpooQy-1 is “Bob”. SpooQy-1 has the firmware with the Kyber-512 implementation uploaded and booted. The ground station has two pieces of software running: the csp-term and our executable implementation of the Kyber API. We use the command scheduler on the satellite to schedule commands that are unknown to the ground station terminal.
\par Two ground stations had been used. One at the NUS Campus in Singapore and one at FHNW in Switzerland. The two ground segments were connected through a quantum-safe version of the Secure Shell Protocol (SSH), which simplified collaborative work, since all ground stations could be remote controlled.

\subsection*{Experiment}
Figure \ref{fig:kex_diagram} shows the performed key exchange where Alice is the initiator. In a first step, one secret/public key pair is generated each for Alice ($sk_{\mathrm{A}}$, $pk_{\mathrm{A}}$) and for Bob ($sk_{\mathrm{B}}$, $pk_{\mathrm{A}}$). Next, the public keys are exchanged between the two stations by downloading/uploading the text files. Alice then starts the key encapsulation mechanism by generating a third key pair ($esk_{\mathrm{A}}$, $epk_{\mathrm{A}}$) which is used as the basis for the common secret key. Note that this key pair is generated using a cryptographically secure RNG. The output of both the key generation ($epk_{\mathrm{A}}$) and start of the authentication ($c_2$) is then uploaded to Bob. Bob then performs several encapsulation and decapsulation operations, which enables him to use several outputs ($K$, $K_1$, $tk'$) to derive the final key. The second output from Bob's encapsulation is then downloaded to Alice again, where a final decapsulation generates the same components for the key derivation as Bob already has. Alice and Bob are now in possession of the same common key. To verify this, we downloaded Bob's key to the ground station to compare it with Alice's key. The full command sequence is shown in the Listing \ref{lst:commands}.

\begin{lstlisting}[caption={command sequence for the KEX experiment},label=lst:commands]
GND > kex-init
SAT > kex_kyb -i
GND > ftp_upload ./PKA.key /flash/PKB.key
GND > ftp_download /flash/PKA.key ./PKB.key
GND > kex-pub -A 
GND > ftp_upload ./ake_senda.txt
SAT > kex_kyb -B 
GND > ftp_download /flash/COMMON.key
GND > mv ./COMMON.key ./SATELLITE.key
GND > ftp_download /flash/ake_sendb.txt
GND > kex-pub -C
GND > diff ./COMMON.key ./SATELLITE.key
\end{lstlisting}

\subsection*{Benchmarking SSH with PQC algorithms}
Since SpooQy-1 reentered Earth right after we performed the key exchange, we could not do benchmarking studies for the RF link. However, we performed tests for a quantum-safe version of the SSH protocol that we used in the Singapore-Switzerland internet link. Here we benchmarked the new NIST round 3 candidates against currently used Elliptic-Curve Diffie-Hellman (ECDH). In Figure \ref{fig:performance_ssh} we show the results for the handshake times in the quantum-safe version of SSH that show the average over 1000 handshakes as a function of different key-exchange algorithms. For all algorithms the authentication was performed with Dilithium 2. It appears that the lattice-based algorithms perform similarly and on par with ECDH. However, the code-based algorithm classic McEliece is taking more time since its public key is substantially larger.

\end{multicols*}

\begin{Figure}
    \centering
    \includegraphics[width=24pc]{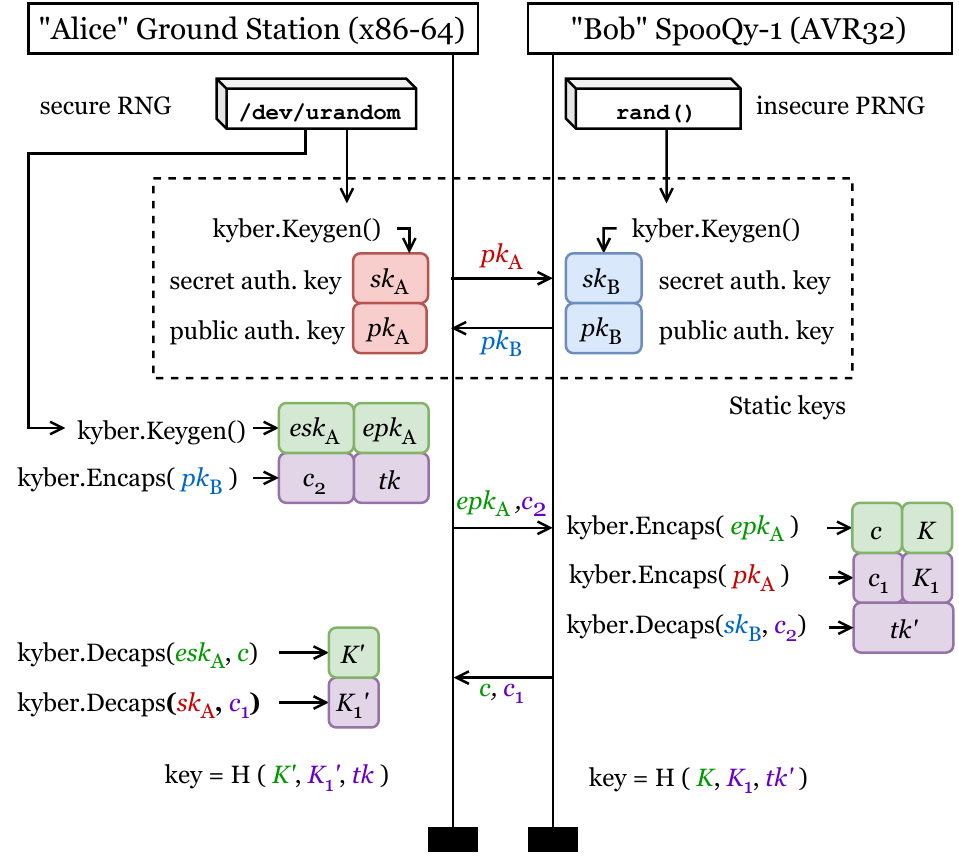}
	\captionof{figure}{Mutually authenticated key exchange where the ground station is the initiator.}
	\label{fig:kex_diagram}
\end{Figure}

\begin{Figure}
    \centering
    \includegraphics[width=34pc]{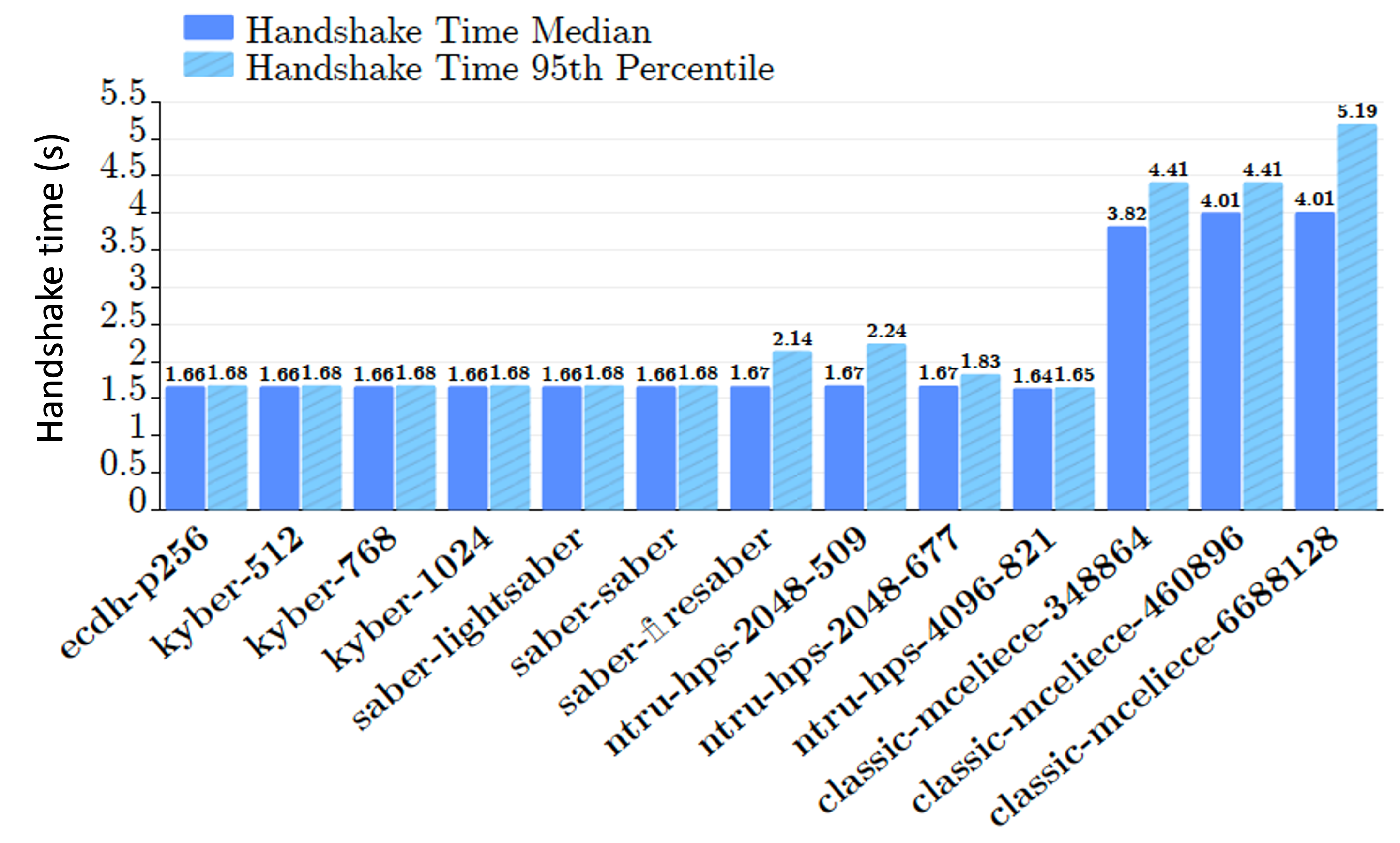}
	\captionof{figure}{Total handshake time in seconds as a function of the round-3 NIST key-exchange algorithms. Median (dark-blue) and 95th percentile (light-blue). The 200 ms round trip time between Singapore and Switzerland is included. The classical key-exchange (ecdh-p256) is shown as a baseline. All key-exchange algorithms are authenticated with Dilithium 2 and the numbers shown are from averaging over 1000 handshake times.}
	\label{fig:performance_ssh}
\end{Figure}

\begin{Figure}
    \centering
    \includegraphics[width=36pc]{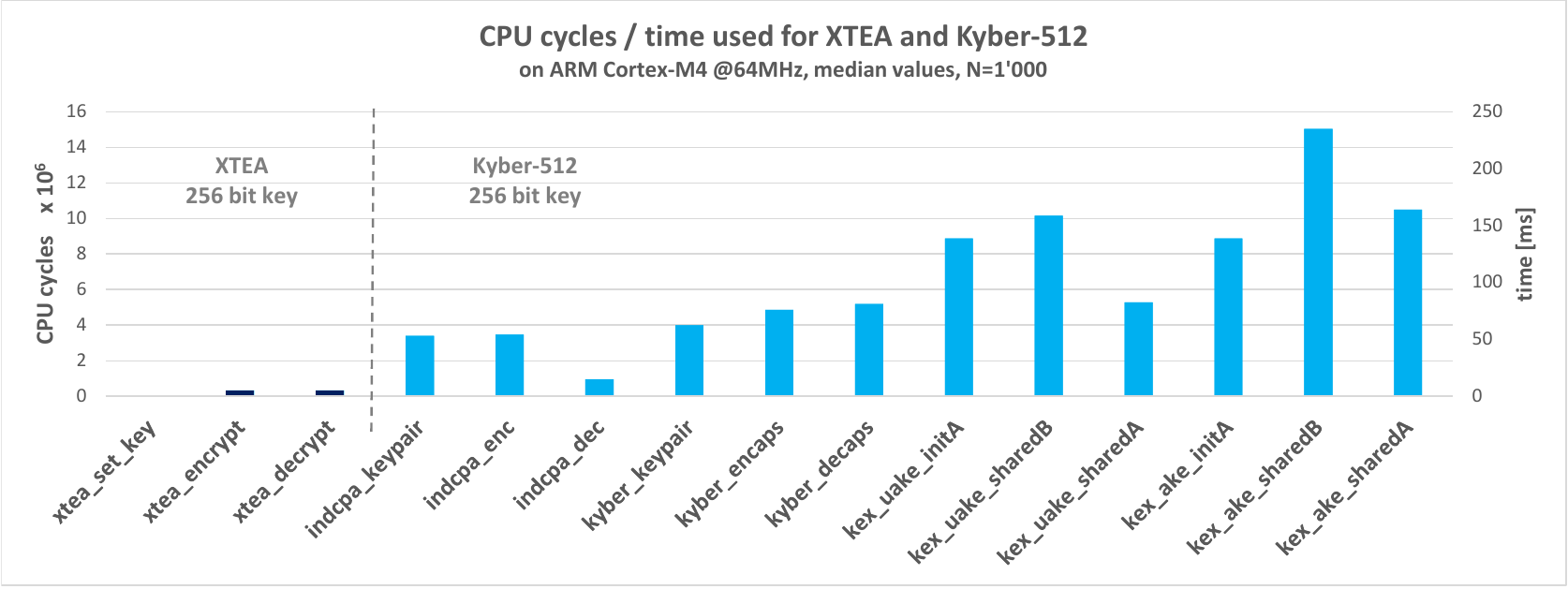}
	\captionof{figure}{Performance measurements on an ARM Cortex-M4 microcontroller.}
	\label{fig:kyber_XTEA}
\end{Figure}

\begin{multicols*}{2}

\par We also conducted benchmarking measurements to compare Kyber-512 to XTEA.
Keep in mind that XTEA is a symmetric block cipher while Kyber is an asymmetric key encapsulation mechanism. The comparison between the two algorithms should therefore only serve as a point of reference for embedded developers.
Since XTEA is the only cryptographic function implemented in libcsp it does make sense to compare its computational effort to the proposed Kyber-512 algorithm.
For the performance measurements we use a hardware that is comparable to SpooQy-1's OBC. Here, we used the STM32F407VG microcontroller based on the ARM Cortex-M4 RISC architecture. Figure \ref{fig:kyber_XTEA} shows the median values for 1’000 iterations of various cryptographic functions executed on this microcontroller.

We see that Kyber-512 needs substantially more resources than XTEA. However, as OBCs on CubeSats become more powerful and the trend is towards using higher performing architectures like the ARM Cortex-A9 (NanoMind Z7000), PQC algorithms should present a feasible alternative to their classical counterparts.

%\section{Special circumstances}

%This demonstration was not planned ahead of time but rather sprung to our minds as an %opportunity to use the remaining time in orbit of SpooQy-1. Within two months’ time we %learned how to implement the Kyber key exchange mechanism into the AVR32 environment. The %development of a proper functioning firmware was further restricted by the circumstance %that the engineering and qualification module (EQM) of SpooQy-1 was non-operational. We %could effectively only debug the code after a lengthy upload procedure to the satellite in %orbit. Nonetheless, we were able to perform the key exchange on the 14th of October during %SpooQy-1's final passes before atmospheric reentry. 

\section*{Conclusion}
We were able to demonstrate a quantum secure key exchange with a nanosatellite in low Earth orbit using the Kyber-512 KEM API. 
Implementing a PQC algorithm on an embedded micro controller brings new weaknesses that need to be addressed by the developer, such as the usage of a cryptographically secure RNG. There is also a potential risk that the implementation of the new algorithms do not support the old MCU architecture currently used for developing nanosatellites. The integration of Kyber into the libcsp project is planned as a follow-up project at FHNW. Since SpooQy-1 has also demonstrated recently a working QRNG \cite{SPOOQY_QRNG}, we may use those random numbers in the next satellite mission SpooQy-2 as a cryptographically secure RNG instead of the PRNGs described previously. For a future project one could take on the task of improving the security features of libcsp by implementing the Kyber algorithm for example. There would however be the challenge of providing a secure RNG which cannot be programmed into the library generically.

\section*{Acknowledgement}
We thank Raffael Anklin for his help with the AVR32 microcontroller, and Frank Imhof for benchmarking the SG-CH connection. 

%\bibliography{bib.bib}

\bibliographystyle{unsrt}

\end{multicols*}
\end{document}